\documentclass[twocolumn,showpacs,preprintnumbers,amsmath,amssymb]{revtex4}

\usepackage{graphicx}
\usepackage{dcolumn}
\usepackage{bm}

\begin{document}

\preprint{APS/123-QED}

\title{Atom reflection echoes and surface matter waves in atom meta-optics}
\author{V. Bocvarski}
\author{J. Baudon}
\email{jacques.baudon@univ-paris13.fr}
\author{M. Hamamda}
\author{M. Boustimi}
\altaffiliation{Present adress : Department of Physics, Umm Al-Qura University, Mekkah, Saudi Arabia}
\author{F. Perales}
\author{G. Dutier}
\author{C. Mainos}
\author{M. Ducloy}
\affiliation{Laboratoire de Physique des Lasers, Universit\'{e} Paris 13, 93430-Villetaneuse, France}

\date{\today}

\begin{abstract}
Evanescent matter-waves produced by an atom wave packet incident onto a repulsive barrier edge can be back-refracted and reconstructed by the application of negative-index ``comoving" potential pulses. One shows that those collapses and revivals of evanescent matter waves give rise to surface matter waves and should be observable via atom reflection echoes issued from the barrier interface. This property, together with the property of inducing negative refraction, makes such potentials the matter-wave counterpart of negative-index materials in light optics.
\end{abstract}

\maketitle
	With the fast development of matter-wave optics, many of the functions previously operated in light optics have been realised: atom diffraction and mirrors, beam splitters, atom lasers, atom holography, etc. [1]. Specific characters of those processes originate in the properties of the associated particle: non-zero atom mass, vacuum dispersion for the ``de Broglie'' waves (implying longitudinal wave packet spreading), scalar character of the atomic wave function, influence of the internal atomic degrees of freedom. Along this viewpoint, novel areas in the field of atom optics are presently explored, including \textit{ e.g.}, the devising of non-diffracting atom nano-beams via a specially designed transverse Stern-Gerlach interferometer [2]. Recently, we have proposed to extend the concept of ``meta-optics'' (known as negative index materials, NIM-s, in light optics [3-4]) to matter waves [5]. The main specificity of NIM ``meta-media'' for atom optics is their transient character, linked to the fact that, for matter waves, the reversal of atom group velocity is operated and can be only transient - contrary to the equivalent process in light optics where the phase velocity is reversed. In [5], it was proposed to use a ``comoving'' potential pulse [6] to reverse the atom group velocity and induce negative refraction. In this letter, we examine the behaviour of the atomic wave packet in negative-index media and analyse its dynamics in classically forbidden regions (evanescent or semi-evanescent matter wave packets). The dynamics of matter waves impinging a static potential barrier is strongly altered by the comoving potential. In particular, we investigate here the possibility of engineering evanescent matter-wave packets and predict the generation of atom reflection echoes as well as surface matter waves at the barrier edge.\\
	It is commonly admitted that, in the semi-classical regime, light and matter waves involving a single direction of space (\textit{e.g}. x), behave identically provided they are monochromatic, \textit{i.e.} provided they involve a single value of the wave number k (or particle's energy). For both kinds of waves, the evanescent wave appears, at a normal incidence with respect to a planar interface, as soon as the refractive index of the second medium is imaginary. This means for light a conducting medium (metal, plasma). For particles, a constant potential energy barrier $V_0$ , extending from x = 0 to x infinite, higher than the initial kinetic energy $E_0$ of the particle, or, in the case of a wave packet, higher than the kinetic-energy distribution, provides an imaginary index since the index for matter waves in the region x $>$ 0 is $n = (1 - V_0/E_0)^{1/2}$. In a 2D geometry (x, z) with the same potential barrier, the motion along z is free, which leads to a total reflection in the $x < 0$ half plane and to an evanescent wave in the $x > 0$ side. The corresponding index is static. However, as shown in [5], a negative index for matter waves is necessarily a transient process, thus needing incident wave packets, \textit{i.e.} some k-momentum distribution. As a consequence, light and matter evanescent waves behave differently as atomic wave packets do not factorize in x and t, due to vacuum dispersion. The existence of evanescent matter waves is one of the fundamental processes introduced by quantum mechanics, and has many outstanding consequences, mainly explored, up to now, for electrons (like electron tunnelling, as in tunnelling microscopy [7], or Josephson Effect in superconductivity [8]). For atomic systems, it has not been much studied, except in molecular spectroscopy [9], collision physics [10] and more recently cold atom physics [11], where evanescent waves are commonly involved (\textit{e.g.} quasi-bound states and resonances of various types). 
	The evolution in x and t of a partially evanescent matter-wave packet, experiencing a pulse of comoving potential, will be described in the following. This comoving potential has the general form:
\begin{equation}
V(t, x) =  s(t)\: \cos \left(2 \pi \frac{x}{\Lambda} \right)
\label{eq1}
\end{equation}
	where s(t) is a signal restricted to a finite time interval (\textit{e.g.} 0, $\tau_1$) and $\Lambda$ is the spatial period. In the following we shall consider, as in previous studies [2,5], magnetic potentials acting on magnetic atoms, such as metastable argon atoms Ar*($^{3}P_2$). Actually any type of potential of the type (1) able to act upon the atomic wave packet motion would be convenient. Similarly, the repulsive potential barrier can be created using a static magnetic field $B_0$ in the half-space $x > 0$ (see figure 1).  

\begin{figure}
\includegraphics[width=6.cm]{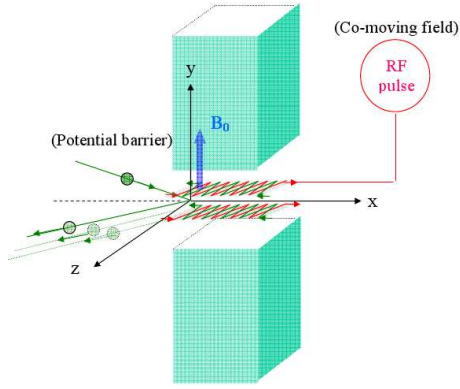}
\caption{Principle scheme of an experiment on evanescent and semi-evanescent wave packets submitted to comoving magnetic potential pulses (see text). A static potential barrier is generated by the static magnetic field $B_0$. Potential pulses, comoving in the x direction, are produced by a generator (``RF pulse''). The incident wave packet propagates in plane (x, z). The reflected wave packet and reflection echoes propagate in a direction of the (x, z) plane, symmetric of the incident direction with respect to the x axis.}
\label{fig.1}
\end{figure}

		We are dealing here with three wave functions, namely  $\Psi_{i} (t, x)$, the incident wave propagating from $x = - \infty $  to x = 0, $\Psi_{r} (t, x)$, the reflected wave propagating backwards from x = 0 to $x = -\infty $, and the transmitted evanescent wave $\Psi(t, x)$ in the region $x > 0$. For a specific value of the wave number k, \textit{i.e.} a specific value E of the total incident energy, the three wave functions have the same time-dependence, in $exp [-i \hbar k^{2} t / (2m)]$. In the absence of comoving field, the conservation of the probability flux leads to the usual transmission and reflection factors (in amplitude): $T(k)= 2 k /(k+i \kappa)$ and $R(k)= T(k)-1$, with $\kappa = \sqrt{a^2 - k^2}$ and $a^2 = (2m/ \hbar^2)V_0 > k^2$. As expected, since the evanescent wave does not carry any probability flux, $\left|R\right| =1$. Similar expressions of R and T are obtained for $k > a$ (partial propagation within the barrier), by replacing $i\kappa$  by $k' = \sqrt{k^2 - a^2} $. In a first approximation, these expressions of R and T factors also hold if the comoving potential is applied at a small but finite distance from the barrier $(0 < x << 1/\kappa )$. In spite of the fact that, -since it does not propagate -, the evanescent wave never really leaves the barrier edge, we shall assume that the effect of the comoving potential pulse V(t, x) can be treated within the only half-space $x > 0$. Under this condition, the time dependent Schr\"{o}dinger equation is:
\begin{equation}
i\hbar~\partial_{t} \Psi =  -\frac{\hbar^{2}}{2m} \partial^{2}_{x} \Psi +\left[V_{0}+V(t,x)\right] \Psi
\label{eq3}
\end{equation}
		where  $\Psi(t, x)$ is the wave function. Let $\Psi_{0}(t, x)$ be the solution in absence of comoving potential, that is (up to a multiplicative constant):
\begin{equation}
\Psi_{0}(t,x) = \int^{+\infty}_{-\infty} T(k) \rho(k) \phi_{0}(t,x,k) dk
\label{eq4}
\end{equation}
		 $\rho(k)$ being the momentum distribution and
\begin{equation}
\phi_{0}(t,x,k) = e^{-\kappa x} e^{-i\hbar k^{2} t / 2m}
\label{eq4a}
\end{equation}
Setting
\begin{equation}
\phi (t, x, k) = f(t, x, k) \phi_{0}(t, x, k)
\label{eq4b}
\end{equation}
		 and assuming that the ``perturbation'' factor f evolves, as a function of x, much slower than $\phi_{0}$, then the term in $\partial^{2}_{x} f$  can be neglected with respect to $\kappa \partial_{x} f$. The validity of this latter approximation will be verified a posteriori. It leads for f to the equation:
\begin{equation}
i\hbar \partial_{t} f =  \frac{\hbar^2}{m} \kappa \partial_{x} f + Vf 
\label{eq5}
\end{equation}
Let us define two new (complex) variables:
\begin{equation}
u =  t + i\frac{m}{\hbar \kappa}~x \ ;\ v =  t - i\frac{m}{\hbar \kappa}~x
\end{equation}

Using (7) it is readily verified that $(\partial_{t} + \frac{i \hbar \kappa}{m}\partial_{x}) = 2 \partial_{v}$. With variables u, v, eq. (6) becomes:
\begin{equation}
\frac{\partial_{v}f}{f} = -\frac{i}{2\hbar}V(u,v)
\end{equation}
Setting $f = A(u) exp[i \varphi(u, v)]$, one gets  $\partial_{v} \varphi = -V/(2 \hbar)$ and finally:
\begin{equation}
\varphi(u,v) = -\frac{1}{2\hbar} \int^{v_{0}}_{v_{1}} dv^{'} V(u,v^{'})
\end{equation}
		 A can be taken as a constant, in so far as its dependence on u is incorporated in the lower bound of the integral. For a comoving potential of the form (1), at a given value of u, setting v' = 2t' - u, one gets from (8):
\begin{equation}
\varphi = -\frac{1}{\hbar} \int^{(t)}_{0} dt^{'} s(t^{'}) \cos\left[\frac{2\pi}{\Lambda}\frac{i\hbar \kappa}{m} (u-t^{'}) \right]
\label{eq6}
\end{equation}
		 where the upper bound is (t) = Min [t,$\tau_1$]. Finally, the phase shift depends on 3 independent variables, k (\textit{via} $\kappa$), t and x (\textit{via} u).\\
	It is interesting to compare this expression of $\varphi$ to that of the phase shift - let us call it  $\varphi_{pro}$ - previously used in the case of a purely propagating matter wave (cf. Ref. [5], Eq. 8): 
\begin{equation}
\varphi_{pro}(k,t)= -\frac{1}{\hbar} \int^{t}_{0} dt^{'} s(t^{'}) \cos\left[\frac{2\pi}{\Lambda}\frac{\hbar k}{m}t^{'} \right]
\label{eq7}
\end{equation}
		 Expression (10) becomes identical to (11) when $i \kappa$ is replaced by k, provided that $\left|u\right|$ can be considered much smaller than  $\left|v\right|$  . In other words, replacing $i \kappa$ by k, this equivalence holds provided that $\left|t - \frac{m}{\hbar k} x \right|<< \left|t+ \frac{m}{\hbar k} x \right|$. This means that, in the case of propagation, all relevant values of t and x are supposed to be close to those related to the propagation of the wave packet centre. Such an approximation is clearly not valid for an evanescent wave packet. The signal s(t) in Eq. (1) is taken as :
\begin{equation}
s(t) = C\epsilon^{2}(t+\epsilon)^{-2}e^{-t/\tau} \ for\ 0\leq t \leq \tau_{1} ~; =0 \ elsewhere
\label{eq8}
\end{equation}
		 where C is a constant. For a magnetic potential $C = g \mu_{B} B_{max} M$, where g is the Land\'{e} factor, $\mu_{B}$ the Bohr magneton, $B_{max}$ the maximum value of the magnetic field magnitude and M the magnetic quantum number. We consider here atoms the spin of which is J = 2, namely argon metastable atoms (Ar* $^{3}P_2$) having a velocity along x axis of 4 m/s (de Broglie wavelength  $\lambda_{dB} = 2.8~nm$), polarized in the M = +2 Zeeman sublevel. The maximum magnitude of the comoving field $B_{max}$ is varied from 0 up to 400 Gauss. The centre of the Gaussian momentum distribution  $\rho(k)$ is $k_0 = 2.244~10^9 m^{-1}$. Evanescent wave packets appear when $k < k_0$ within a repulsive potential barrier of momentum height $a = k_0$ (this corresponds to a magnetic field of 127 Gauss). This value of the height has been chosen in order to get a ``semi-propagative'' regime of the wave packet. The standard deviation is  $\delta k = 0.005~k_0$. In s(t), time parameters are different from those previously used since they must be adapted to typical characteristics of the wave packets in the region $x > 0$. In the present case we take $\epsilon = 7.4~\mu s, \tau = 0.37�~\mu s, \tau_1 = 0.60~\mu s$. The spatial period is $\Lambda = 2~\mu m$ [12]. The effect of the comoving pulse on the wave function in half space $x > 0$ is simply obtained by incorporating the factor $exp(i \varphi)$ into the expansion of this wave function over k (Eq. 3). At this point, let us examine the validity of approximation used previously to get the simplified equation (6). Actually it is rather poor for the original comoving potential in $cos(2 \pi x/\Lambda )$, in the sense that the ratio $\left|\kappa~\partial_{x} f \right|/ \left|\partial^{2}_{x} f \right|$, which is zero for $\kappa = 0$ $(k = k_0)$, becomes larger than 1 only when $ \left|k - k_0\right| > \delta k$. The reason is that such a potential starts abruptly at $x = 0$. The situation is greatly improved by slightly shifting the spatial dependence, into, for instance, $cos (2 \pi x /\Lambda - \pi / 10) $ which is quite easy from an experimental point of view. In such a case the ratio is larger than 10 as soon as $\left|k - k_0\right| > \delta k / 20$. In other words, only a narrow central slice of the spectrum violates the approximation. It is expected - and it has been verified - to be of little importance in so far as, for the spatial dependence of the wave packets, it involves large distances $(x > 20/ \delta k = 1.57~\mu m)$. In what follows, the spatial dependence in $cos (2 \pi x /\Lambda - \pi / 10) $ has been adopted, at least when semi-evanescent wave packets are considered. For fully evanescent ones (\textit{i.e.} $a = 1.02~k_0$), this correction has not to be used in so far as the questionable part of the spectrum ($\kappa \approx 0$) corresponds to a very low intensity.
		 
\begin{figure}
\includegraphics[width=8.5cm]{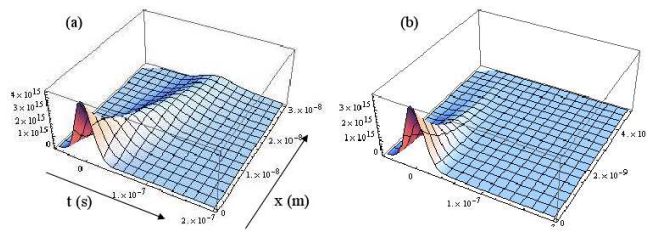}
\caption{Profile in t and x of the squared modulus $\left|\psi_0\right|^2$ of the unperturbed semi-evanescent wave function. (a) The height a (in momentum) of the potential barrier is chosen equal to central value $k_0$ of the incident momentum distribution. At x = 0, the time dependence is Gaussian. For $x > 0$, one observes as expected the decay characteristic of the evanescent part together with a partial wave packet propagating at a positive group velocity and broadening with time. (b) Same as (a) for a pure evanescent wave ($a = 1.02~k_0$). No more propagation is observed: exponential spatial decay of the wave at t = 0; wave-packet fully vanishing for $t>0$.}
\label{fig.2}
\end{figure}
	 		 
		 The profile in t and x of the squared modulus $\left|\psi_0\right|^2$ of the unperturbed evanescent wave is shown in figure 2. Figures 3 a-b show the squared modulus $\left|\psi\right|^2$ of the wave function modified by the comoving field, for $B_{max} = 50$ and 28 Gauss. Two effects are clearly seen in these figures: first, a negative refraction effect bringing the wave packet backwards to the barrier edge after some delay; simultaneously, a significant enhancement of the wave on its way back to the barrier edge. As the spatial range of the comoving potential is infinite, one could think the reflected part of the wave packet is also brought back to the barrier edge by the negative refraction. Actually this does not occur because the central momentum $k_0$ in the region $x < 0$ is much higher than any propagating momenta within the barrier, the order of which is $\delta k$. Consequently the magnitudes of $B_{max}$ considered here ($B_{max} =~50~Gauss$) are not sufficient to bring the reflected part back on the barrier edge. Therefore the revival observed at $x = 0^+$ after some delay, originates from the matter wave inside the barrier. Because of the boundary continuity conditions at x = 0, this revival generates a secondary wave packet, an atom ``reflection echo'' in the region $x < 0$. Actually similar echoes are obtained even when the barrier height is such that the propagative part is largely suppressed, \textit{e.g.} $a = 1.02~k_0 = k_0 + 4~\delta k$. In such a case (see fig. 4a), no appreciable internal negative refraction effect is seen (\textit{i.e.}, no propagation effect inside the potential barrier), whereas the wave echo remains. Note the narrowing of the echo (the width of which is similar to the initial one), produced by a time-reversal property of the meta-medium [13]. On the other hand, in the case of an almost pure propagation, \textit{e.g.} for $a = 0.99~k_0$, the negative refraction inside the barrier, accompanied by the narrowing of the wave packet [13], becomes the main phenomenon. Reflected and transmitted intensities emitted from the barrier are characterized by internal reflection and transmission factors given by expressions similar to the previous ones. The internal reflection generates an additional wave packet which propagates towards positive x as soon as the comoving pulse is over. A similar enhancement effect is observed when a second pulse, delayed by $0.6~\mu s$, is applied (see fig. 3c). This is a clear evidence for the beginning of a temporal series of collapses and revivals of the semi-evanescent matter wave at the potential barrier. The same type of result is obtained with a pure evanescent wave packet (see fig. 4b). Successive revivals of evanescent waves are also predicted when a single comoving potential pulse of sufficient duration is applied (fig. 4c). This behaviour of the wave packet when subsequent pulses of comoving potential are applied strongly suggests the existence of a surface matter wave, which remains confined along the barrier edge (x = 0 axis) while exhibiting a series of temporal rebounds along the t axis. Because of the boundary conditions, each revival of the evanescent wave at the barrier generates in free space a new, retarded, reflected matter wave packet propagating in the $x < 0$ direction, an atom reflection echo which should be observable.  This should be a signature of the behaviour of the evanescent wave packet inside the barrier. Another observation mode would consist in cutting off the potential barrier at some distance, smaller than the evanescent decay length, and measure the transmitted intensities.
\begin{figure}
\includegraphics[width=8.cm]{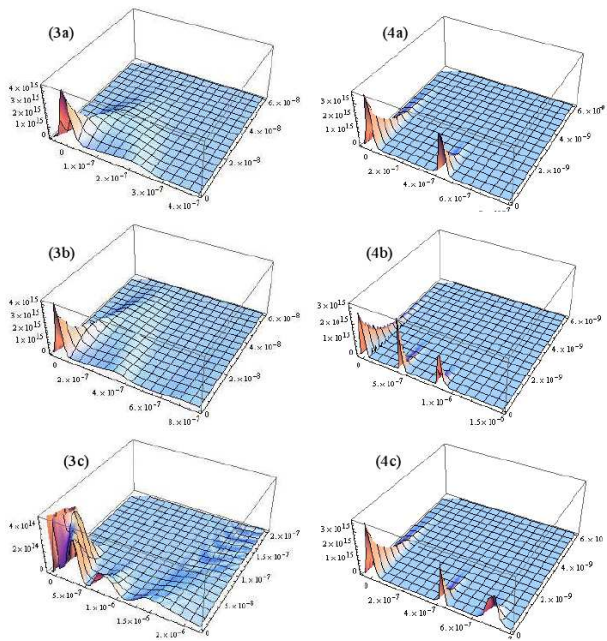}
\caption{ (Left side) Same as fig. 2a (semi-propagating wave packet, $a = k_0$), in the presence of a single comoving potential pulse of duration $\tau_1 = 500~ns$ and magnitude (see text) (a) $B_{max} = 50~Gauss$, (b) $B_{max} = 28~Gauss$; in (c), two identical subsequent pulses ($B_{max} = 28~ Gauss$) are applied, the second one being delayed by 600 ns with respect to the first one. Note that the vertical scale is magnified by 10. Sequential rebounds are seen. \\
FIG. 4: (Right side) Same as fig. 3, with a barrier height $a = 1.02~k_0$ (pure evanescent wave). The comoving potential pulse has a magnitude $B_{max} = 35~Gauss$. (a) Single comoving pulse of duration $\tau_1 = 425~ns$; (b) two sequential pulses are applied, within a time delay of 425 ns. This makes appear a second echo. (c) Evanescent wave revivals with a single potential pulse of duration 500 ns.}
\label{fig.34}
\end{figure}
		 
		 To summarize, adequate pulsed comoving potentials have proven to be able not solely to induce a negative refraction of matter waves, but also to cause drastic effects on the dynamics of evanescent or semi-evanescent matter waves, generating repeated enhancements in the vicinity of the barrier edge, which gives rise to surface matter wave, as well as atom reflection echoes. This makes these ``negative index media'' a counterpart of left-handed media in light optics, with however important differences related to the different natures of matter and light waves. One should emphasize the close relationship between the present approach (\emph{i.e.} interaction of virtual atoms of the evanescent matter-wave with ``real'' magnetic or light fields) and that demonstrated long time ago about the interaction of ``virtual'' photons with real atoms, like saturated absorption in evanescent light waves [14]. Various applications can be imagined for matter waves : some of them are similar to those of left-handed materials in light optics, such as atom wave focusing (meta lenses and atom nano-lithography), cloaking, etc.; some others are specific of matter waves such as beam splitters and interferometers and guided matter wave along a potential barrier edge.


Authors are members of the \textit{Institut Francilien de Recherche sur les Atomes Froids} (IFRAF).


\begin{thebibliography}{99}
\bibitem{spatial} Such a short spatial period might be difficult to implement with a magnetic field. 
       A different approach should be to use a near-resonant quasi-standing light wave.
\bibitem{narrowing} Narrowing of the matter-wave packet in negative-index media is a general feature of 
       atom ``meta-optics'', which is connected to its time-reversing properties. It will be 
       presented in detail in a forthcoming publication.
\end{thebibliography}
\end{document}